# Copper phosphate micro-flowers coated with indocyanine green and iron oxide nanoparticles for *in vivo* localization optoacoustic tomography and magnetic actuation

Daniil Nozdriukhin[1,2], Shuxin Lyu[1,2,3], Jerome Bonvin[1,2], Michael Reiss[1,2], Daniel Razansky[1,2] and Xosé Luís Deán-Ben[1,2]

[1] Institute for Biomedical Engineering and Institute of Pharmacology and Toxicology, Faculty of Medicine, University of Zürich, Winterthurerstrasse 190, Zurich 8057, Switzerland

[2] Institute for Biomedical Engineering, Department of Information Technology and Electrical Engineering, ETH Zürich, Rämistrasse 101, Zurich 8093, Switzerland

[3] Department of Medical Imaging, Shanxi Medical University, Taiyuan 030001, China

## Abstract

Efficient drug delivery is a major challenge in modern medicine and pharmaceutical research. Micrometer-scale robots have recently been proposed as a promising venue to amplify precision of drug administration. Remotely controlled microrobots sufficiently small to navigate through microvascular networks can reach any part of the human body, yet real-time tracking is crucial for providing precise guidance and verifying successful arrival at the target. *In vivo* deep-tissue monitoring of individual microrobots is currently hampered by the lack of sensitivity and/or spatio-temporal resolution of commonly used clinical imaging modalities. We synthesized biocompatible and biodegradable copper phosphate micro-flowers loaded with indocyanine green and iron oxide nanoparticles to enable *in vivo* individual detection with localization optoacoustic tomography. We demonstrate magnetic actuation and optoacoustic tracking of these decorated micro-flowers at a per-particle level. Functional super-resolution imaging achieved via tracking intravenously injected particles provides a means of identifying microvascular targets and quantifying blood flow, while the versatile carrying capacity can be further exploited for transporting multiple types of drug formulations.

## Introduction

Drugs and contrast agents reach the target site in the body through intricate microvascular networks of the circulatory system [1–3]. Advances in nanomedicine have resulted in increasingly complex nanoparticles and nanocarriers leading to significant theranostic advances [4–6]. However, the targeting efficiency of passive accumulation or chemical binding to specific receptors remains relatively low, even if particles are directly injected into the bloodstream [7,8]. As a representative example, a thorough literature review revealed that only approximately 0.7% of the administered nanoparticles are delivered to solid tumors [9,10]. Thereby, pharmaceutical research has placed significant emphasis on establishing the optimal blood concentration and circulation time to achieve therapeutic or diagnostic purposes while minimizing the risk of overdose and toxic effects [11]. In recent years, micrometer-scale robots have emerged as a new means of transporting chemical substances throughout the body that can redefine the precision and effectiveness of drug administration. Microrobots remotely controlled by external acoustic or electromagnetic forces can be sufficiently small to flow through microvascular networks and reach the region of interest without causing capillary arrest, thus massively improving the transport of their cargo to virtually any part of the body [12–17].

The effective implementation of microrobotic delivery of drugs and other substances faces two main challenges, namely 1) identification of the precise location where therapeutic or diagnostic measures are needed and 2) precise guidance of the microrobots to reach the designated target site(s). Biomedical imaging modalities are poised to play a pivotal role in overcoming these key obstacles. However, existing clinical imaging systems have a limited capacity to distinguish specific biological indicators (biomarkers) of disease, while *in vivo* imaging and tracking of individual microparticles have also remained elusive. Particularly, whole-body scanners based on magnetic resonance imaging (MRI), x-ray computed tomography (CT), or positron emission tomography (PET) have insufficient spatial resolution to distinguish microvascular alterations characteristic of many pathological conditions and insufficient temporal resolution to visualize small objects flowing in blood. The high temporal resolution of ultrafast ultrasound has been exploited to track individual microbubbles and provide super-resolution angiographic images of tissue cross-sections [18,19] and entire volumes [20–22]. However, detection of other types of microparticles with a reduced acoustic mismatch is hampered by the substantially weaker pulse-echo responses of these. Localization optoacoustic tomography (LOT) has recently been proposed as an alternative approach for volumetric super-resolution angiography that further enables mapping of oxygen saturation within microvascular networks [23]. More importantly, LOT provides a significantly higher level of flexibility than ultrasound for microparticle tracking as virtually any particle encapsulating a sufficiently absorbing substance can potentially be detected.

Petal-shaped particles exhibit an exceptionally high surface-to-volume ratio allowing for an unparalleled potential for the transportation of functional materials [24–28]. Particularly, so-called copper nano- to micro-flowers can be readily synthesized from commonplace materials such as copper sulphate as a source of copper ions, phosphate-buffered-saline (PBS) as a source of phosphate ions, and a template consisting of proteins or polysaccharides [29–31]. They exhibit a highly porous three-dimensional hierarchical structure enabling a high loading capacity e.g. via layer-by-layer assembly of a multilayer coating, a method proven to be a viable option for producing silica-core particles providing LOT contrast on a per-particle basis [32]. Herein, we propose layer-by-layer coating of copper micro-flowers (further referred to as copper roses or CuR) for the development of magnetically steerable microrobots that can be individually detected and tracked with LOT. For this, highly-absorbing indocyanine green (ICG) and paramagnetic iron oxide particles (FeNP) were embedded. LOT tracking of particles is shown to result in super-resolution images of microvascular networks within optically opaque tissues that potentially help define functional biomarkers to establish the target site. Magnetic control of the particles is demonstrated in microfluidic chips *in vitro* and in the mouse ear *in*

*vivo*. Also, they are shown to be biodegradable and no major adverse effects were observed in mice for two weeks following *in vivo* intravenous injection. The high flexibility of the proposed microrobotic system to further incorporate multiple types of molecules and nanoparticulate formulations is poised to offer an unprecedented level of precision for the controlled delivery of therapeutic drugs.

# Results

### Loading of copper micro-flowers

The envisioned microrobotic platform, aimed at transporting substances throughout the body, relies on magnetically sensitive copper phosphate micro-flowers also providing optoacoustic (OA) contrast. Magnetic actuation was enabled with FeNP loaded on the extensive surface of the CuR, which together with an additional coating of ICG further results in strong ultrasound responses generated via absorption of light. Microparticle synthesis was based on $Cu_3(PO4)_2$ nucleation on chains of dextran sulfate sodium salt (DxS) in phosphate buffered saline (PBS) followed by layer-by-layer deposition of the load (Fig. 1a, see methods for a detailed description). Specifically, the pristine particles were first coated with a linking layer of poly-(diallyl-dimethylammonium chloride) (PDDA) to induce a positive surface charge. This was followed by the deposition of four consecutive bilayers of FeNP/poly-L-ornithine (PLO) and ICG/PLO, and an eventual poly-(sodium-styrene sulfonate) (PSS) layer preventing fouling in biological media. Zeta-potential measurements performed after each step demonstrated the expected charge switch after application of opposite-charged layers (Fig. 1b). Substantial differences between the particle morphology of pristine and coated particles were also observed in scanning electron microscopy (SEM) images of these (Figs. 1c and 1d, see methods for details). Specifically, the petals of the coated particles showed a much rougher and thicker surface associated with FeNP deposition on the polyelectrolyte cushion (Fig. 1d). This enhanced roughness was consistent across the entire microparticle surface, validating that the intended high chemical payload was achieved. The composition of the pristine (CuR) and coated (CuIF) particles was analyzed with energy-dispersive X-ray spectroscopy (EDS, Fig. 1e). Peaks corresponding to sulphur, copper, oxygen, phosphorus, chlorine, and carbon, observed in both CuR and CuIF, are related to the materials used in the microparticle synthesis procedure (PBS, $CuSO_4$, and DxS). Additional peaks corresponding to iron and nitrogen were observed after the coating procedure, indicating FeNP and ICG deposition, while the observed increased intensity of carbon and sulfur peaks is related to the polyelectrolyte chains. According to SEM images, the measured particle diameter was 4.2±0.2 µm.

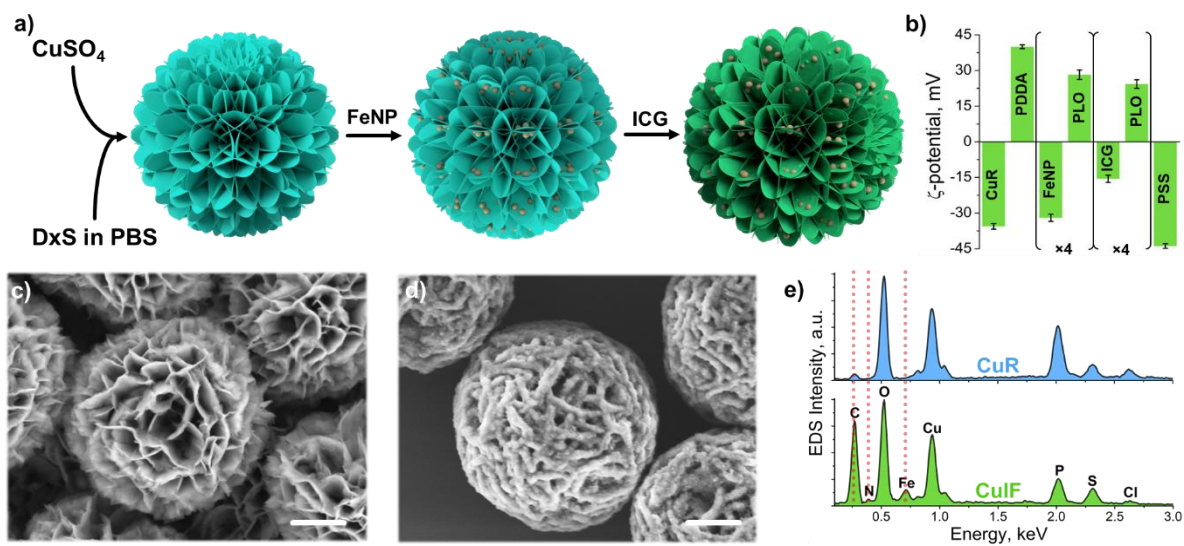

**Figure 1:** a) Schematic description of the synthesis and coating of copper micro-flowers. First, $CuSO_4$ and Dextran Sulfate (DxS) were mixed in PBS to obtain the core material (CuR). Subsequently, the CuR structures were coated by a layer-by-layer assembly consisting of consecutive layers of iron oxide nanoparticles (FeNP) and indocyanine green (ICG). b) Evolution of zeta-potential of the particles considering the last adsorbed surface layer. c) Scanning electron microscopy (SEM) image of the uncoated CuR. d) SEM image of the FeNP/ICG-coated CuR (CuIF). Scalebars – 1 um. e) Energy-dispersive X-ray spectroscopy (EDS) spectra of pristine (CuR) and coated (CuIF) micro-flowers. Red dotted lines indicate the peaks increasing after deposition of polyelectrolyte cushion, ICG and FeNP.

**Biosafety and biodegradability**

The biosafety and potential toxic effects of the proposed CuIF were assessed both *in vitro* and *in vivo*. Different concentrations of CuIF ($10^7$-$10^{10}$ particles per mL) were added to Chinese hamster ovary (CHO) cultured cells and co-incubated for 24 h. An alamarBlue cell viability assay was then performed, which quantifies cell metabolism based on 590 nm fluorescence emission for 540 nm excitation wavelength (see methods for details). No significant reduction of cell metabolism was observed for most of the tested concentrations (Fig. 2a), even for cells uptaking particles identified in brightfield (Fig. 2b) and confocal microscopy (Fig. 2c) images. Internalization of microcarriers by target cells can play an essential role in maximizing drug efficacy provided the particles degrade at a controlled rate without causing harm to other cells. The biodegradation of CuR and CuIF was assessed by monitoring the CHO cultured cells for 24 h with brightfield microscopy. Complete disassembly of CuR and partial discoloration of CuIF were observed after 24h incubation (Fig. 2d). The first substantiates the biodegradability of the core particles, while the second is ascribed to slow shell decomposition and cargo (ICG) release. These findings validate the potential of CuIF as effective drug microcarriers. The low toxicity of CuIF was further confirmed *in vivo* with two groups of mice (n=4 per group) intravenously injected with 100 uL of PBS (control group) and CuIF (study group). No significant weight loss, physical or behavioral changes were detected for two weeks following administration of CuIF (Fig. 2e). Also, no significant differences in hematological parameters

(measured at days 1, 7, and 14 post injection) or biochemical parameters (measured at day 14 post injection) were observed between control and study groups. Particularly, similar levels of alanine aminotransferase, an enzyme often used as a marker indicating damage to liver cells, and creatinine, a waste product signaling impaired kidney function, were observed between both groups.

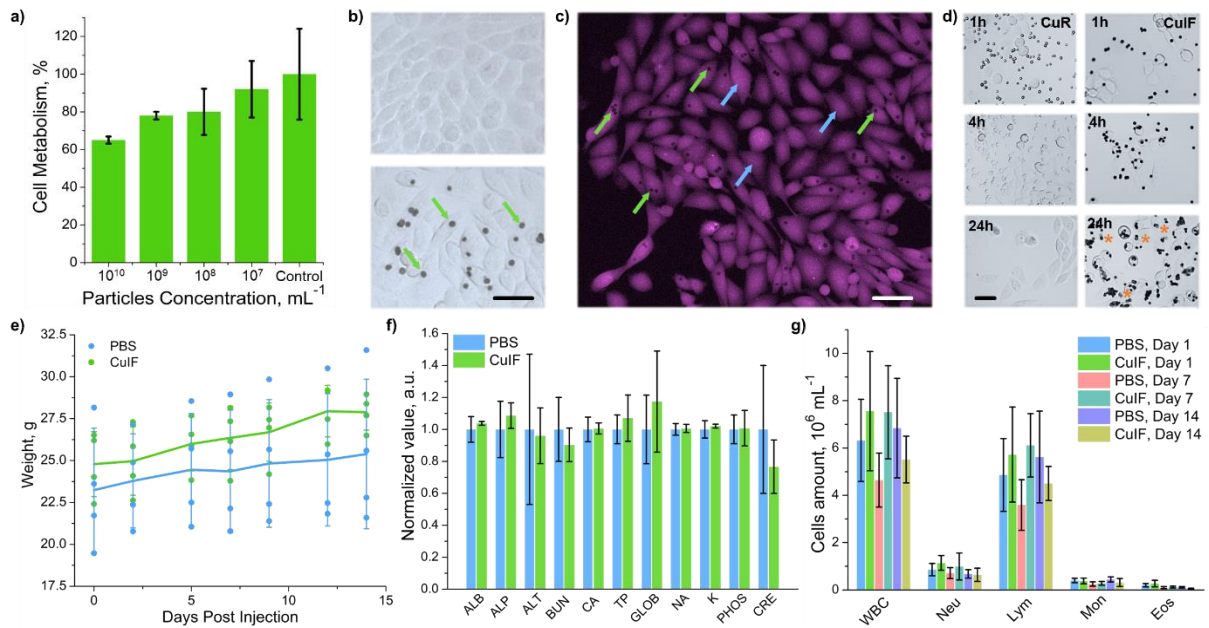

**Figure 2:** a) alamarBlue cell viability assay in Chinese hamster ovary (CHO) cells incubated with different concentrations of FeNP/ICG-coated micro-flowers (CuIF). b) Brightfield image of the CHO cells without (top) and with (bottom) internalized particles. Green arrows point to particles taken by cells. Scalebar -20 µm. c) Confocal microscopy image of CHO cells with alamarBlue inside the cell cytoplasm incubated with CuIF for 24 h. Green arrows point to particles inside the cells (dark spots) and blue arrows point to empty cells. Scalebar – 20 µm. d) Biodegradation of the pristine (CuR) and CuIF particles in the presence of CHO cells. Scalebar – 20 µm. e) Mouse weight monitoring in two groups (n=4) of mice injected with PBS and CuIF. f) Blood biochemistry taken at day 14 post-injection of PBS and CuIF. ALB – albumin, ALP - alkaline phosphatase, ALT - alanine transaminase, BUN – blood urea nitrogen, CA – calcium, TP – total protein, GLOB – globulin, NA – sodium, K – potassium, PHOS – phosphates, CRE – creatinine. Values are normalized to average values for the control (PBS) group (ALB – 4.48±0.36 g/dL, ALP - 107.25±18.84 U/L, ALT – 30.25 ± 14.22 U/L, BUN – 15.5 ± 3.11 mg/dL, CA – 12.53 ± 0.97 mg/dL, TP – 5.45 ± 0.49 g/dL, GLOB – 0.98 ± 0.21 g/dL, NA – 154.75 ± 5.56 mmol/L, K – 8.28 ± 0.45 mmol/L, PHOS – 14.03 ± 1.26 mg/dL, CRE – 0.3 ± 0.12 mg/dL). g) Blood haematology taken at days 1, 7, and 14 post-injection of PBS and CuIF. WBC – total amount of white blood cells, Neu – neutrophils, Lym – lymphocytes, Mon – monocytes, Eos – eosinophils.

**Super-resolution optoacoustic imaging**

OA contrast is generated by absorption of light, thus represents a versatile means of detecting multiple types of molecules and particles of different sizes [33,34]. OA imaging is generally performed with optical wavelengths in the so-called near-infrared (NIR, 650-1400 nm) window to maximize the penetration of light within biological tissues, with reported depths up to several centimeters [35]. The absorption spectra of the synthesized particles were first characterized with ultraviolet (UV)-NIR optical spectroscopy (Figs. 3a and 3b). This showed NIR absorption of ICG and copper sulphate on the pristine particles, UV absorption of FeNP, and almost no absorption of the DxS solution in PBS. The UV-NIR spectrum of CuIF was slightly red-shifted with respect that of ICG, arguably indicating aggregation of the dye molecules on the polyelectrolytes or their oxidation on FeNP. OA spectroscopic measurements performed with a tunable laser source were shown to match UV-NIR spectroscopy readings (Fig. 3b, see methods for a detailed description), excluding the potential influence of scattering and/or fluorescence of the particles in the measurements. The high per-particle absorption of the CuIF particles further favors their use as contrast agents for LOT. LOT enables angiographic imaging beyond the acoustic diffraction barrier via detection and tracking of particles in the broodstream [23,32,36]. The feasibility of *in vivo* LOT imaging of the murine brain was assessed in a mouse intravenously injected with a bolus of a CuIF suspension (100 uL, $10^8$ particles per mL). The particles could be volumetrically tracked in a sequence of OA images (800 nm excitation, 100 frames per second) acquired with a spherical array (Fig. 3c). The LOT image built with the localized positions of particles in 30000 frames (300 sec) clearly enhances the resolution and angiographic information of OA images acquired with the same array (Figs. 3d and 3e). Additionally, the high temporal resolution achieved with the volumetric OA imaging system allows for quantification of blood flow velocity via particle tracking (Fig. 3f). Taken together, these unique imaging capabilities can greatly assist in identifying therapeutic target regions and mapping the transport of microcarriers through microvascular structures, leveraging blood flow to facilitate magnetic steering and guidance.

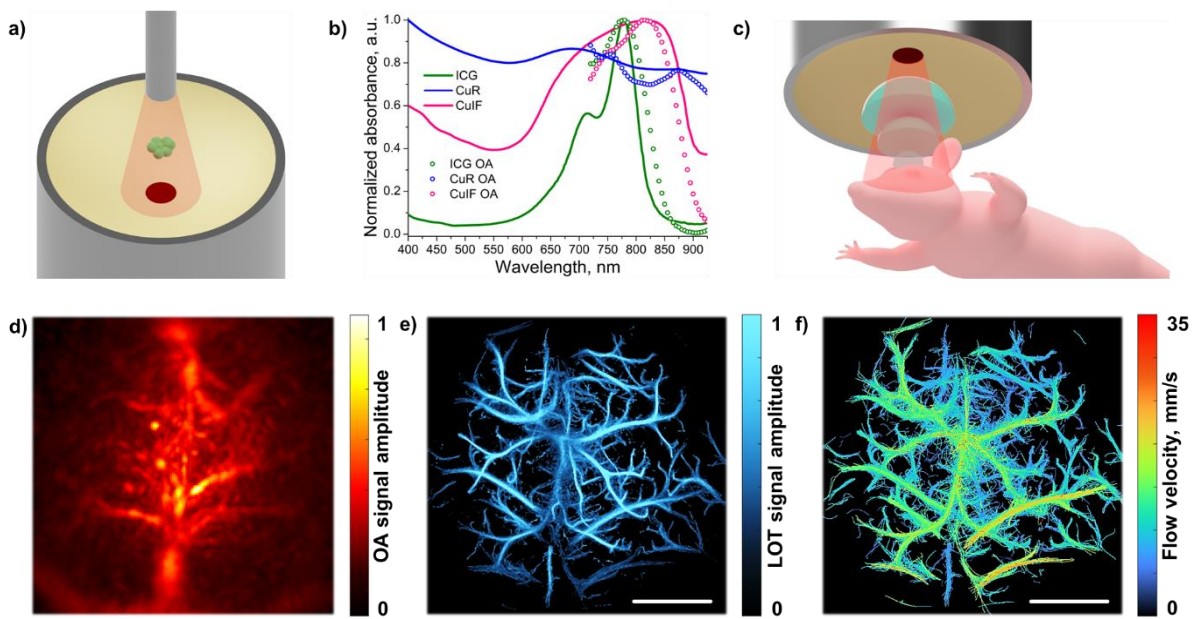

**Figure 3.** a) Schematic representation of the system used for optoacoustic (OA) spectroscopy. b) UV-NIR spectra (solid lines) and OA spectra (circles) of indocyanine green (ICG) and the synthesized pristine (CuR) and FeNP/ICG-coated (CuIF) micro-flowers. c) Schematic representation of the system used for

OA and localization optoacoustic tomography (LOT) imaging of the mouse brain. d) OA image of the mouse brain. e) Corresponding LOT image. f) Flow velocity map of the same region. Scalebars – 5 mm.

**Magnetic control of particles and cells**

The magnetic properties of the particles and the feasibility of magnetic guidance of these were investigated in microfluidic experiments (Fig. 4a). First, the behavior of the CuIF particles in a static aqueous medium (no flow) was investigated. The weak Brownian motion of the particles, when no magnetic field was present, caused them to disperse and become approximately equally distributed across the microfluidic channel (Fig. 4b). After activation of the magnet, the particles started to move towards the magnetic pole with a noticeable chaining behaviour, typical for nano- and micro-particles at relatively high concentrations [37,38] (Fig. 4c). Subsequent removal of the external field led to a rapid disorganization of the particles (within 20 sec), indicating almost superparamagnetic properties. Considering that CuIF can be internalized by cells, cell manipulation and control can also be achieved. This was demonstrated with a murine macrophage culture co-incubated with the CuIF microparticles for 24 h to ensure high cellular uptake (Figs. 4d and 4e). Towing of cells with particles attached to the surface or internalized was possible, while chaining of particles inside and outside cells resulted in a "tugboat" effect enabling cell tethering (Fig. 4f). This can set the stage for new approaches in single-cell studies with cell spheroids, macrophage manipulation, or other applications [39–42].

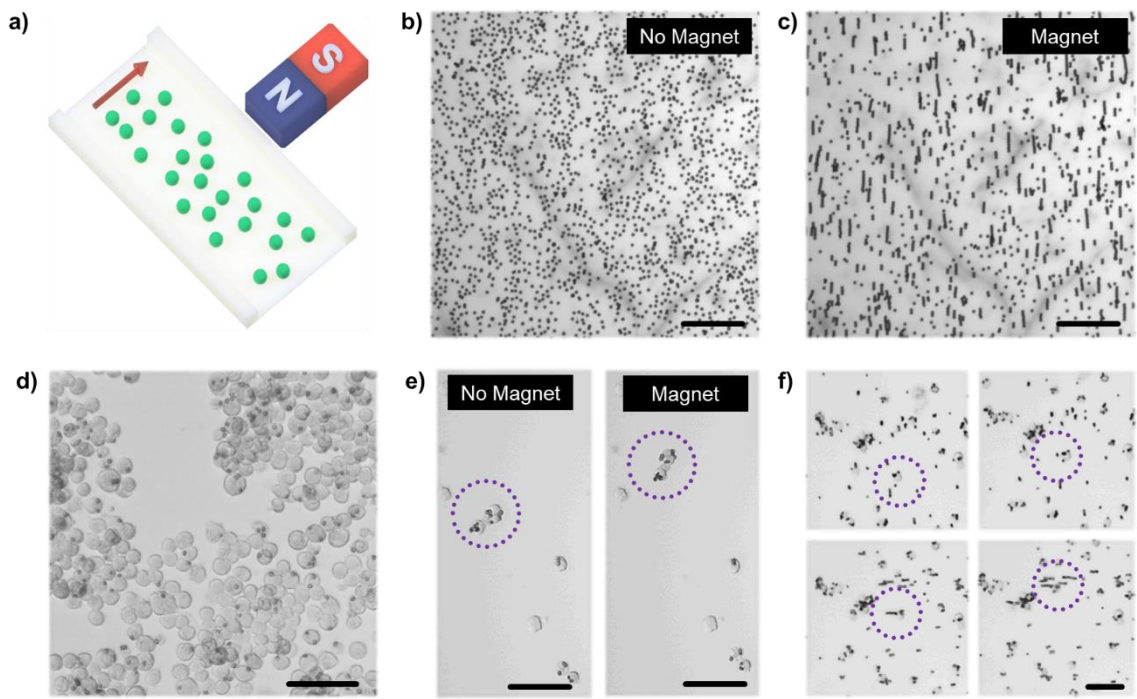

**Figure 4.** a) Schematic representation of the magnetophoresis experiments, where particles in a microfluidic capillary are moved by a permanent neodymium magnet. b) Brightfield optical image of the FeNP/ICG-coated micro-flowers (CuIF) inside the microfluidic chip without application of an external magnetic field. c) Brightfield optical image of the CuIF microparticles inside the microfluidic chip following application of an external magnetic field. Chaining behavior of the CuIF microparticles is observed. d) Brightfield optical image of murine macrophages with internalized CuIF microparticles (dark spots inside cells). e) Magnetic tow of the cells with internalized particles. f) CuIF microparticles chaining with particles inside the cell and further towing of the cell inside the magnetic field.

**Magnetic actuation and optoacoustic guidance *in vivo***

The feasibility of *in vivo* trapping and controlling the motion of CuIF microparticles in the bloodstream under OA guidance was demonstrated in the mouse ear. Volumetric (three-dimensional) imaging of the mouse ear was achieved with a spherical array pointing upwards with the enclosed volume filled with agar (Fig. 5a). The mouse was placed in supine position with the ear coupled with ultrasound gel to the agar block. The OA image revealed major vessels in the mouse ear, while CuIF microparticle tracking following intravenous injection of a 100 uL bolus of a particle suspension served to identify the flow direction in the visible vessels (Fig. 5b). Singular value decomposition (SVD) filtering applied to a sequence of OA images allowed for separation of the slowly varying background signal from ear vessels from the quickly moving spheroids associated with CuIF microparticles in the bloodstream (Fig. 5c). A change in behavior of CuIF particles was observed after an electromagnet mounted over the ear was turned on (dashed green circle in Fig. 5d). Specifically, some of the particles were clustered near the pole of the electromagnet, with a particle being shown to be pulled in a direction opposite to the blood flow (Fig. 5d).

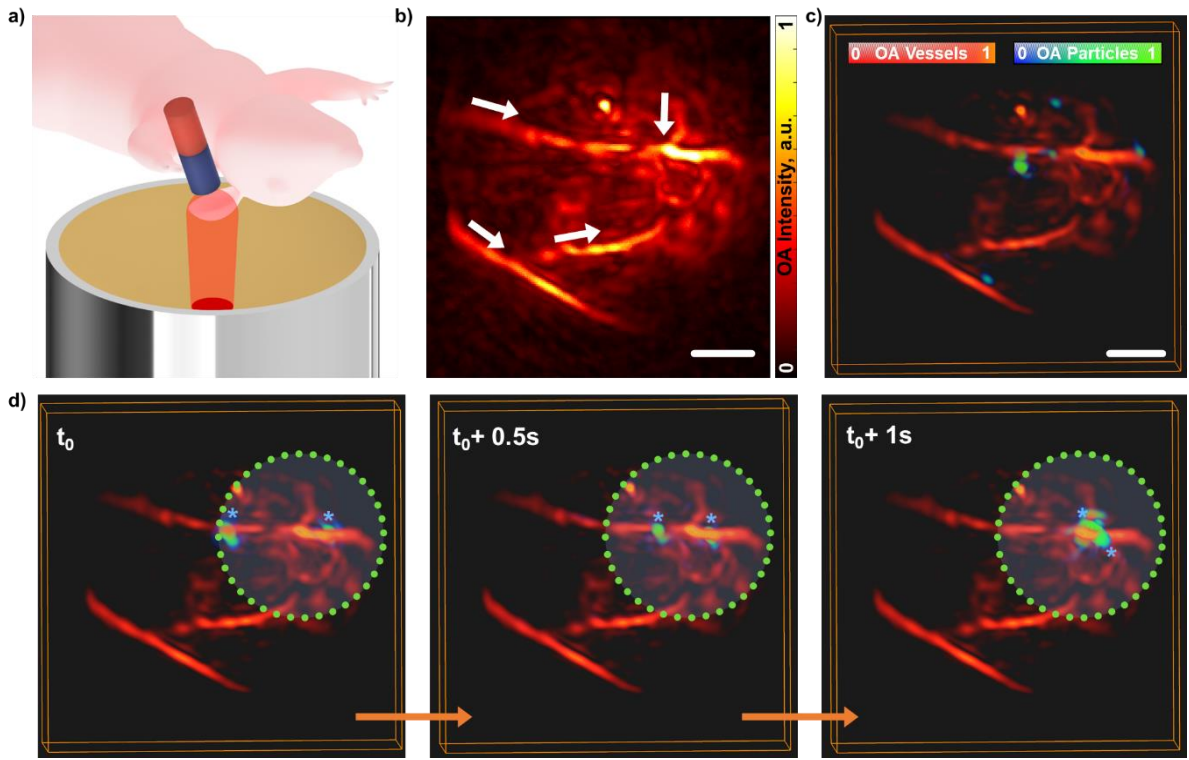

**Figure 5.** a) Schematic representation of the *in vivo* magnetic guiding experiment. b) OA image of the murine pinna vasculature near the scapha. White arrows indicate the flow direction in the vessels. Scalebar – 1 mm. c) OA image of the same region after administration of CuIF microparticles along with the singular value decomposition (SVD)-based filtered image showing individual CuIF particles (green dots). Scalebar – 1 mm. d) Actuation on two CuIF particles induced by an external magnetic field pulling them towards each other. The particle in the right is shown to move against the blood flow.

## Discussion and conclusions

Versatile adherence of chemicals benefits from the porous three-dimensional structure of the synthesized particles, characterized by a surface-to-volume ratio notably higher than alternative microparticulate configurations. The size and morphology can be additionally controlled by alternating the pH level and content of reactants in the solution to achieve a broad variety of sizes in a range of 0.1-10 micrometers [43–47]. The exceptional flexibility offered by layer-by-layer coating of these types of particles provides a unique means of achieving new types of formulations both in terms of chemical payload and functional properties. The high-loading capacity of the proposed microparticles is thus expected to result in a unique ability to carry and deliver theranostic substances such as drugs or imaging contrast agents. Also essential is the demonstrated biocompatibility and biodegradability. CuIF particles have been shown to undergo controlled degradation over time, which is crucial to facilitate elimination from the body and minimize potential enduring adverse effects [48]. Incorporation of clinically approved substances such as ICG and FeNP is not expected to significantly increase long-term toxicity so the suggested microparticles show great potential for extensive preclinical investigations and also hold promise for an eventual clinical translation.

A high load of contrast media is needed to enable accurate monitoring of the location of the particles and to ensure that they reach the target tissue. Also important is the fact that individual tracking of small microparticles as they move throughout the bloodstream demands spatio-temporal resolution ranges unreachable with most biomedical imaging modalities. Arguably, OA represents the most suitable approach for *in vivo* particle tracking due to its high sensitivity, versatile contrast mechanism, relatively deep penetration, and high-frame-rate imaging capacity [49–56]. As shown herein, OA visualization of individual microparticles sufficiently small to traverse capillary networks can further be exploited to render super-resolution LOT images of deep-seated microvascular networks. The amount of CuIF particles intravenously injected to produce super-resolved images of the brain microvasculature was considerably higher than that of microdroplets recently used to demonstrate the *in vivo* feasibility of LOT [23]. As a consequence, detailed microangiographic images could be rendered for a relatively low number of acquired frames, which effectively results in an increased temporal resolution of LOT. Indeed, new types of microparticles are essential to optimize the LOT imaging performance, mainly associated to the per-particle OA signal intensity and the maximum concentration of particles that can be safely injected in the bloodstream. LOT provides an exceptional functional imaging ability, allowing for simultaneous mapping of blood flow and oxygen saturation at a resolution beyond the acoustic diffraction limit across entire three-dimensional areas [23,57,58]. This facilitates quantification of microvascular alterations in the target tissue. OA guidance and tracking of the proposed microrobots can then massively impact disease diagnosis and controlled drug delivery.

Magnetic actuation of microrobots enables precise remote control over their motion and position, thus allowing navigation within the body with great accuracy [17,59–61]. This offers significant advantages to alternative chemical or thermal actuation methods generally resulting in uncontrolled motion [62,63]. Also, external power delivery enables continuous actuation without relying on internally stored energy within the particles, which is essential to reach remote areas. Microrobotic actuation can alternatively be achieved via acoustic forces sharing many of the advantages of magnetic fields [13,64,65]. However, magnetic fields are more stable across different tissue types and can interact with multiple types of particles, while ultrasound waves can interfere with OA signals. Thereby, OA monitoring of magnetic actuation appears to represent an optimal combination for microrobotic intravascular guidance. Also important is the fact that the synthesis procedure can easily be scaled to produce a large amount of particles. Recently, manufacturing of functional microstructured materials of similar size has been enabled with additive manufacturing methods, which are however impractical to synthesize a sufficient number of particles for imaging and therapeutic purposes [66,67].

Encapsulation of drugs within the CuIF particles can easily be achieved with the proposed layer-by-layer approach without compromising the OA or magnetic properties. The observed magnetic responsiveness *in vivo* emphasizes the potential clinical relevance, paving the way for non-invasive, precise drug delivery strategies. However, technical and regulatory challenges must be overcome before a potential clinical translation. Of particular importance is defining a strategy for the release of the microparticle cargo to extravascular targets. Extravasation can be facilitated if the size of the particles is reduced and sufficiently strong forces are applied. Individual OA tracking of smaller particles is however yet to be demonstrated. On the other hand, ejection or rupture of the microparticle coating has been achieved with sufficiently strong laser pulses [68–71]. This may be explored as a viable method for drug release, particularly considering that microparticle ablation can be monitored with OA.

In conclusion, the demonstrated *in vivo* magnetic actuation coupled with super-resolution OA tracking of biocompatible microflowers is poised to result in new methods for precise identification of microvascular targets within the body and efficient administration of theranostic substances. These unprecedented capabilities anticipate a groundbreaking shift in the field of drug delivery, potentially opening a new era of precision medicine.

## Materials and methods

### Materials

Copper sulfate pentahydrate ($CuSO_4 \times 5H_2O$), dextran sulfate sodium salt from Leuconostoc spp. (DxS, Mw ~500 kDa), Dulbecco-modified phosphate-buffered saline (PBS) solution (DPBS, without calcium chloride and magnesium chloride), indocyanine green powder (ICG), poly-l-ornithine-hydrobromide (PLO, Mw > 100 kDa), poly-(diallyl-dimethylammonium chloride) solution (PDDA, 20%, Mw ~200 kDa), sodium chloride (NaCl), poly-(sodium-styrene sulfonate) (PSS, 70 kDa) were purchased from Sigma Aldrich. Citric acid-capped superparamagnetic iron oxide nanoparticles (FeNP) with 9±2 nm diameter were procured from TetraQuant Ltd. Dulbecco's Modified Eagle's Medium (DMEM), alamarBlue cell assay kit, fetal bovine serum (FBS) and 96-well plates were purchased from Thermo Fisher Scientific Inc. J774A.1 murine macrophages and Chinese Hamster Ovarian (CHO) cells were obtained from CLS Cell Lines Service GmbH. Double-deionized water (DI) water with an electrical resistivity of 18.2 MOhm·cm produced by a Millipore Milli-Q A10 system was used in all experiments.

### Synthesis of copper phosphate micro-flowers

Copper phosphate hybrid micro-flowers (also referred to as copper roses or CuR) were synthesized according to the modified and upscaled protocol from [31]. First, 120 mM $CuSO_4$ in DI water and 1 mg/ml DxS in PBS solutions were prepared. Then, 2 ml of the $CuSO_4$ solution was rapidly added to 10 mL of the DxS solution in a 15 mL tube, well-shaken for 1 min and left on the tube rack for 60 min. After the first 15 min of reaction, the light-blue mixture turned to a turbid-blue colour indicating $Cu_3(PO_4)_2$ nucleation on DxS chains. After 1 h, the reaction was stopped by 5 times centrifugation at 200 rcf for 1 min and rinsing with DI water. The resulting suspension of CPM was further used for layer-by-layer deposition.

### Surface coating

The surface of the synthesized CuR was coated with ICG and FeNP using a layer-by-layer approach. First, both PSS and poly-L-ornithine were dissolved at 1 mg/ml concentration in 0.5 M NaCl aqueous solution. ICG was prepared at 1 mg/ml in DI water. The layers were deposited for 15 min each followed by 3 1-min washing cycles at 200 rcf. Specifically, layers of PLO, FeNP, and ICG were subsequently deposited to produce a CPM/PDDA/[FeNP/PLO]$_4$/[ICG/PLO]$_4$/PSS structure (CPM-FeICG). The particle coating steps were controlled by UV-Vis spectroscopy and zeta potential measurements. UV-Vis spectra were measured using spectrophotometer Infinite Pro 200, Tecan. The ICG-containing samples were diluted 20 times, and the samples without the dye were measured at the original concentration. The results were normalized to a maximum of each spectrum. ζ-potential measurements were made on the Zetasizer Nano ZS90, Malvern Panalytical. All samples were diluted 10 times in deionized water and placed in a folded capillary cell (DTS1070, Malvern). Results were processed with Zetasizer software 8.00. Each measurement was made at 24°C and repeated three times.

**Scanning electron microscopy**

Scanning electron microscopy (SEM, Hitachi SU5000) was used to assess the morphology and composition of the synthesized particles. First, pre-cleaved p-type silicon chips were consecutively cleaned with acetone, ethanol and water using an ultrasonic bath. Then, the samples were drop-cased on the silicon chips and dried at room temperature. SEM microphotographs were collected at 3 kV accelerating voltage. Ultim Max 100 detectors (Oxford Instruments) mounted at 90° to each other were used for energy dispersive spectroscopy analysis at 10 kV accelerating voltage to prove the element composition of the particles processed in AZtec software.

*In vitro* and *in vivo* toxicity assays

The biocompatibility of the microparticles was assessed with an alamarBlue assay carried out on a Chinese hamster ovary (CHO) cell culture. First, cells were cultivated in DMEM supplemented with 10% FBS in a humidified incubator at 37°C in 5% $CO_2$. After that, a 96-well plate was seeded with $8×10^3$ cells per well, and filled with 200 μL of cell medium without Phenol Red. 20 uL suspensions of particles of different concentrations (from $10^7$ to $10^{10}$ particles per mL) were added to each row of the well plate (10 wells per row), leaving one row as the control corresponding to cells fed with PBS. After 24 h incubation, cells were washed with PBS and 20 μL of alamarBlue stock solution was added according to the standard protocol to each well. Cells were then additionally left for incubation for 3 h. After that, 100 μL of the cell-processed medium was transferred to a fresh 96-well plate to form a reading replica without the influence of cells and residual particles inside them, which was then loaded into the well-plate reader (Infinite M200, Tecan). The fluorescence of the medium was measured at 540 nm excitation and 590 nm emission wavelengths. The cellular uptake of particles and cell viability were also verified with brightfield optical microscopy (Primostar 3, Zeiss) and confocal optical microscopy images with fluorescence signal from alamarBlue (LSM 800 inverted microscope, Zeiss; 561 nm laser, 600 nm long-pass emission filter). To check the biodegradability of the microparticles, cell cultures were monitored with the brightfield microscope for 24 h.

The biosafety of the synthesized microparticles was assessed in female Swiss mice (n = 8, 7 weeks old). These were split into control (n = 4) and experimental (n = 4) groups injected i.v. with a single dose of 100 μL of saline and the microparticle suspension ($2×10^9$ particles per mL), respectively. Animals were randomly assigned to the two groups. Mice were scored and weighted at days 1, 2, 5, 7, 9, 12 and 14 following i.v. administration and subsequently sacrificed. Haematological analysis of blood samples was performed by using a BC5000-Vet analyzer (Mindray) at days 1, 7 and 14 post-injection. Clinical biochemistry was performed in a VetScan VS2 analyzer (Zoetis) with the collected serum samples at the final time point. The biosafety study was done following Spanish and European regulations and approved by the Xunta de Galicia.

**Optoacoustic spectroscopy**

The optoacoustic (OA) performance of the particles was first evaluated with OA spectroscopy measurements. For this, the volume enclosed with a 5 MHz central frequency, 512-element spherical array transducer [72] (Imasonic SaS) pointing upwards was filled with 1% agar gel to provide acoustic coupling. Then, a 4 µL droplet containing a suspension of microparticles was deposited on a cover glass slide located on top of the agar gel. A tunable fibre-bundle coupled nanosecond optical parametric oscillator (OPO) laser (EVO II, InnoLas GmbH) was used to illuminate the sample from the top and excite OA signals in the wavelength range 700-1000 nm with a 5 nm interval. The acquired data was analysed using Matlab 2022b to extract the wavelength-dependent signal amplitude.

**Magnetic guidance assay**

A suspension of particles was injected into a single-capillary microfluidic chip (5 mm capillary width, 800 um capillary height, Ibidi µ-Slide I) mounted on a brightfield upright optical microscope (Primostar 3, Zeiss) equipped with a CMOS camera (Ace 2, Basler). A permanent neodymium magnet (40x10x10 $mm^3$,) was placed at the border of the chip. The induced motion of the particles was recorded and quantified using optical flow calculations. To further assess the magnetic performance of the synthesized microparticles, a 40/60 w/w glycerol-in-water solution was prepared as a liquid mimicking the viscosity of blood. The same microfluidic chip was used to demonstrate particle deflection by a magnet in the flow generated by a syringe pump (up to 2.3 mm/s, Aladdin Syringe ONE, World Precision Instruments). A macrophage cell culture with internalized microparticles was further used as a model object to demonstrate the particle towing capabilities inside chips containing a static cell medium (Ibidi µ-Slide I, 5x0.8 $mm^2$ channel cross-section) and a moving cell medium (up to 16 mm/s, Fluidic 144, Microfluidic Chip Shop; 100x100 $um^2$ channel cross-section) mimicking the capillary blood flow.

**Animal Imaging Experiments**

Two female athymic nude mice (~21 g body weight, 8 weeks old) were used for OA imaging and tracking of the synthesized particles *in vivo*. For this, the animals were anaesthetized with isoflurane (5% v/v for induction and 1.5% for maintenance, Abbott, Cham, Switzerland) in an oxygen/air mixture (100/400 mL/min). The mice were housed in ventilated cages inside a temperature-controlled (21±1°C) and humidity-controlled (55±10%) room under an inverted 12 h dark/light cycle. Pelleted food and water were provided *ad libitum*. All experiments were performed under the Swiss Federal Act on Animal Protection and were approved by the Cantonal Veterinary Office Zürich.

***In vivo* optoacoustic tomographic imaging**

OA tomographic imaging of the brain vasculature of the first mouse was performed. The eyes were covered with dexpanthenol cream, and the head was fixed in a stereotactic frame. The mouse was placed in a prone position on a heating pad covered with soft tissue to maintain a constant body temperature (37°C). A 100 µL suspension of the synthesized particles in PBS at an approximate concentration of $10^8$ particles per mL was injected through the tail vein. OA images were captured with a 7 MHz central frequency 512-element spherical array transducer (Imasonic SaS, Voray, France) at 800 nm wavelength and 100 Hz pulse repetition frequency for 5 min (30000 frames): 30 s before injection, 30 s during injection, and 4 min after injection. The laser beam was guided through a 3 mm core liquid light guide with NA ~0.55 inserted into the transducer array central cavity.

**Localization optoacoustic tomography**

Localization optoacoustic tomography (LOT) images were reconstructed from the datasets referred to above according to the procedure described in [23,32]. In brief, the raw signals were first bandpass filtered with cut-off frequencies 0.1-7 MHz. Then, the tissue clutter and high-frequency noise were removed by applying a single-value decomposition (SVD) filter. For this, the Casorati matrices of subsets with dimensions (493x512)x200 were considered to filter the eigenvectors in the range of 25-130, corresponding to the signals from fast-moving microparticles. The local intensity maxima were detected in the filtered datasets and then a tracking algorithm (simpletracker.m available on MathWorks ©Jean-Yves Tinevez, 2019, wrapping Matlab Munkres algorithm implementation of ©Yi Cao 2009) was considered to estimate the displacement of the microparticles for consecutive frames. A maximal linking distance of 0.5 mm was selected, which corresponds to a maximum particle velocity of 50 mm/s for the imaging frame rate.

**In vivo optoacoustic monitoring of intravascular magnetic guidance**

The ability to trap and manipulate the synthesized particles *in vivo* was evaluated via OA tomographic imaging of the ear of the second mouse. For this, the animal was mounted on a supine position on top of 7 MHz 512-element spherical array transducer pointing upwards with the enclosed volume filled with agar. The ear was coupled to the imaging system using ultrasound gel and the images were acquired at 100 Hz repetition rate and 800 nm excitation wavelength for 4 min (24000 frames): 30 s before injection, 30 s during injection, and 3 min after injection. The holding electromagnet with 400 N pulling force (-MS-5030-24VDC, Intertec Components ITS) was mounted over the ear and turned on 1 min after administration of the microparticles. The acquired data were reconstructed and analyzed using Matlab 2022b.

## Acknowledgements

X.L.D.B. acknowledges support from the Helmut Horten Stiftung (project Deep Skin). D. R. acknowledges support from the Swiss National Science Foundation (310030_192757) and the US National Institutes of Health (R01-NS126102-01). The authors gratefully acknowledge ScopeM for their support and assistance in this work.